\documentclass[reprint,superscriptaddress,amsmath,amssymb,aps,prb,longbibliography]{revtex4-2}
\usepackage[utf8]{inputenc}
\usepackage{graphicx}
\usepackage{dcolumn}% Align table columns on decimal point
\usepackage{bm}% bold math
\usepackage[colorlinks,citecolor=blue,linkcolor=blue,urlcolor=blue]{hyperref}
\usepackage[normalem]{ulem}

\begin{document}

\title{$d_{x^2-y^2}$-wave Bose Metal induced by the next-nearest-neighbor hopping $t^{\prime}$}

\author{Zhangkai Cao}
\thanks{These authors contributed equally.}
\affiliation{School of Science, Harbin Institute of Technology, Shenzhen, 518055, China}

\author{Jianyu Li}
\thanks{These authors contributed equally.}
\affiliation{School of Science, Harbin Institute of Technology, Shenzhen, 518055, China}
\affiliation{Shenzhen Key Laboratory of Advanced Functional Carbon Materials Research and Comprehensive Application, Shenzhen 518055, China.}
 
\author{Jiahao Su}
\affiliation{School of Science, Harbin Institute of Technology, Shenzhen, 518055, China}
\affiliation{Shenzhen Key Laboratory of Advanced Functional Carbon Materials Research and Comprehensive Application, Shenzhen 518055, China.}

\author{Tao Ying}
\affiliation{School of Physics, Harbin Institute of Technology, Harbin 150001, China}
 
\author{Ho-Kin Tang}
\email{denghaojian@hit.edu.cn}
\affiliation{School of Science, Harbin Institute of Technology, Shenzhen, 518055, China}
\affiliation{Shenzhen Key Laboratory of Advanced Functional Carbon Materials Research and Comprehensive Application, Shenzhen 518055, China.}

\date{\today}

\begin{abstract}

Superconductivity arises when electrons form Cooper pairs with phase coherence. In contrast, a lack of phase coherence in Cooper pairs can lead to an uncondensed metallic ground state known as the Bose metal state. In this study, we investigate an attractively interacting fermionic system with nearest-neighbor (NN) hopping (t) and next-nearest-neighbor (NNN) hopping ($t^{\prime}$) anisotropy between two species of spins in a two-dimensional (2D) lattice. Utilizing the constrained path quantum Monte Carlo (CPQMC) method, we demonstrate the existence of a $d_{x^2-y^2}$-wave Cooper pair Bose metal (CPBM) phase with $t^{\prime}/t > 0.7$. The CPBM phase exhibits a dome-like structure in the phase diagram of filling $n\sim0.65$, with the maximal region around an optimal $t^{\prime}/t \sim 0.2$, suggesting that an appropriate value of $t^{\prime}$ facilitates the formation of the Bose metal. Furthermore, we find that a Bose metal formed by fermions with a closed Fermi surface confirms that the crucial condition for this exotic phenomenon is primarily the anisotropy of the Fermi surface, rather than its topology. Our finding of the $d_{x^2-y^2}$-wave CPBM demonstrates the same pairing symmetry as the pseudogap behavior in cuprates, and its experimental realization in ultracold atom systems is also feasible. 

\end{abstract} 
\maketitle

The cuprate superconductors \cite{Bednorz1986} have sparked significant interest over the past three decades not only due to their high-temperature ($T_c$) unconventional superconductivity. The peculiar behavior of the pseudogap (PG) phenomena in high-$T_c$ superconductors, which might be closely related to the microscopic mechanism of superconductivity, still awaits a well-recognized explanation \cite{Lee2006-qq,Keimer2015-jp,Singh2021-nh}. One of the most puzzling fact is that the onset of PG phase is accompanied with a gap with ``$d_{x^2-y^2}$-wave like" symmetry opening below the characteristic temperature ($T^*$) with no superconductivity signal~ \cite{hashimoto2014energy}.

Various types of order are proposed to explain the $d$-wave-like PG behavior. In the famous resonating valence bond (RVB) theory \cite{Anderson1987-cv,Zhang1988-dw}, the quantum spin liquid (QSL) is argued as the origin of the PG phase, where the charge and spin degrees of freedom are separated, defines a PG of spinons below $T^*$ when holon condensation is absent \cite{PhysRevB.71.064502}. As an alternation of the QSL phase, the Bose liquid or Bose metal is also propsed to explain the strange metal behavior in PG phase, which presumes that Cooper pairs are dominant charge carriers for the electric transport not only in the superconducting (SC) but also in the metallic phases, constituting a conducting quantum fluid instead of a superfluid \cite{Phillips2003-dd}.

In other words, the existence of a Bose metal phase, which Cooper pairs instead of electrons as the primary charge carriers is a bosonic system \cite{Yang2022-ml}, could potentially explain the strange metal state in high-$T_c$ superconductors. 
Indeed, from Bardeen-Cooper-Schrieffer (BCS) to Bose-Einstein condensation (BEC) crossover theory \cite{Chen_undated-qc}, incoherent Cooper pairs begin to form at temperatures $T_c < T < T^*$, before they Bose condense and superconduct at $T_c$, which is at the heart of BCS-BEC crossover theory.  

Bose metal is argued to leave fingerprint in microscopic model of hard-core bosons with ring exchange on multileg ladders, which called bosonic $J-K$ model \cite{Block2011-fj,Mishmash2011-wn}. As a variant of Bose metals, Cooper pair Bose metal (CPBM) has been theoretically proposed to exist in 2D systems with no polarization \cite{Feiguin2009-nu,Feiguin2011-ib}. The anisotropic spin-dependent Fermi surface plus attractive interactions lead to an effective model of Cooper pairs with a ring-exchange term, that may allow to realize a paired but non-superfluid Bose metal phase. The Cooper pairs would form a collective state with gapless excitations along a Bose surface but no condensate in momentum space.

Notably, the fingerprint of CPBM phase in fermions system has been observed in the qausi-1D systems like two-leg ladder \cite{Feiguin2011-ib} and four-leg ladder \cite{Block2011-is}. In our recent work \cite{Cao2024-qd}, we have found the CPBM phase in the 2D $t-U$ Hubbard model, with onsite attraction interaction ($U$) and the nearest-neighbor (NN) spin-dependent hopping ($t$) anisotropy, in which fermions are paired as bosons, and the uncondensed Cooper pairs form a non-superfluid Bose metal phase. But the boson correlation of Bose metal is mainly the $d_{xy}$-wave, induced by $t$. In a previous study \cite{Jiang2013-io}, it was suggested that adding next-nearest-neighbor (NNN) hopping ($t^{\prime}$) anisotropy might induce the $d_{x^2-y^2}$-wave correlation in the Bose metal phase.

The NNN hopping $t^{\prime}$ is also argued to strongly affect the properties of superconductivity and other orders, as has been revealed in the previous studies in the Hubbard model \cite{Huang2001-lf,Wachtel2017-ai,yang2020quantum,Jiang2019-km,The_Simons_Collaboration_on_the_Many-Electron_Problem2020-ko,Hirayama2018-tm,Hirayama2019-qt}.
A delicate interplay between superconductivity and density wave orders tunable via NNN hopping $t^{\prime}$ is revealed by using extensive density matrix renormalization group (DMRG) studies of the $t-t^{\prime}-U$ Hubbard model at hole doping concentration $\delta = 0.125$ on four-leg cylinders \cite{Jiang2019-km}. The semi-classical Monte Carlo calculations show that introduction of $t^{\prime}$ results in a finite temperature PG phase that separates the small $U$ Fermi liquid and large $U$ Mott insulator.  
We have discovered the presence of Bose metal phase in $t-U$ Hubbard model \cite{Cao2024-qd}, but the influence of the NNN hopping $t^{\prime}$ on the ground state properties of Bose metal phase remains an outstanding theoretical question. 

Here, we choose to focus on the $t-t^{\prime}-U$ Hubbard model, utilizing the constrained path quantum Monte Carlo~(CPQMC) \cite{Zhang1995-hn,Zhang1997-mr}, to look at the effect of $t^{\prime}$ on the Bose metal phase. Our key conclusions are: (i) We found the exotic CPBM phase in $t-t^{\prime}-U$ Hubbard model, emerging when a spin-dependent anisotropy suppresses the ordinary $s$-wave pairing, and emphasize that the appropriate value of $t^{\prime}$ facilitates the formation of the Bose metal phase; (ii) our analysis identifies primary competition between $d_{xy}$-wave and two types of $d_{x^2-y^2}$-wave within the CPBM phase, when $t^{\prime}$ is large enough ($t^{\prime}/t > 0.7$), the system becomes dominated by $d_{x^2-y^2}$-wave Bose metal phase; (iii) we find that a Bose metal formed by fermions with a closed Fermi surface confirms that the crucial condition for this exotic phenomenon is primarily the anisotropy of the Fermi surface, rather than its topology. This $d_{x^2-y^2}$-wave Bose metal phase with uncondensed bosons in 2D, provides a controllable route to enhance CPBM and control the closeness of the Fermi surface, offering insights into the theoretical understanding of the pseudogap phase.

\begin{figure}[b!]
    \centering 
    \includegraphics[width=1\linewidth]{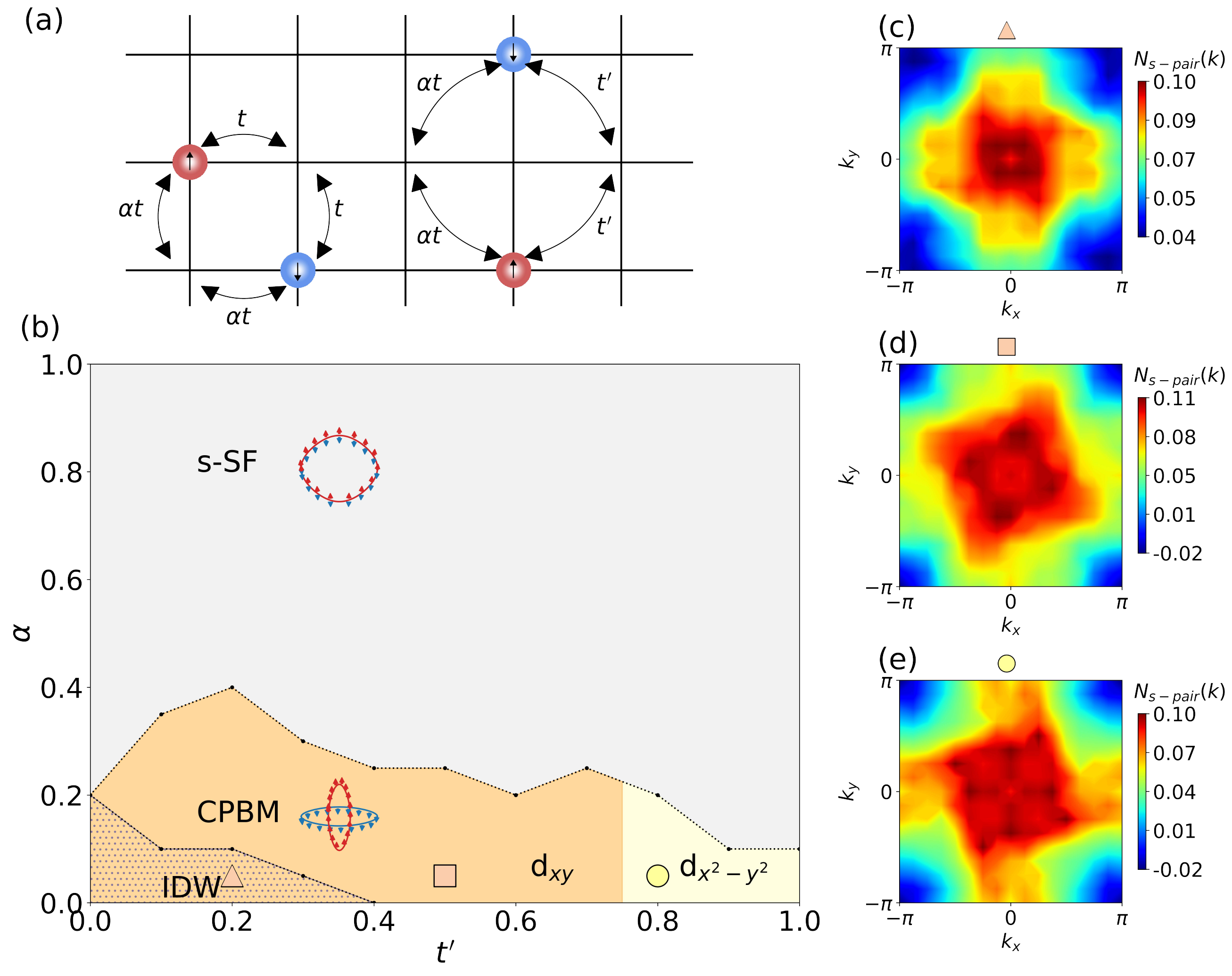}
    \caption{(Color online) Model and phase diagram. (a) Illustration of the model with spin-dependent anisotropic hopping amplitudes on a square lattice. (b) Schematic zero-temperature phase diagram at $U= -3$ (the unit is $t$) for electron filling $n \sim 0.65$, exotic Cooper pair Bose metal (CPBM) exists in the regime with strong spin-dependent anisotropy that suppresses the ordinary $s$-wave superfluid ($s$-SF). IDW is an incommensurate density wave that exists in strong anisotropy and small $t^{\prime}$. CPBM phase divided into two types with $t^{\prime} \sim 0.7$ as the boundary, and dominated by $d_{xy}$-wave and $d_{x^2-y^2}$-wave correlation between bosons, respectively. The inset shows the illustration for conventional $s$-wave pairing by up and down spins, and nonzero momentum $s$-wave pairing by up and down spins. (c)-(e) CPQMC simulation result of the $s$-wave pair momentum distribution function $N_{\mathrm s-pair}({\bf k})$ for different $t^{\prime} = 0.2,\ 0.5,\ 0.8$ at $\alpha =0.05$ with electron filling $n \sim 0.65$ on 16 $\times$ 16 lattice, showing obvious existence of nonzero momentum Bose surface, which is the signal for determining the exotic CPBM phase. The determination of each phase region in the phase diagram is inferred from our CPQMC data (details provided in Supplemental Material \cite{Cao2024t-supp}). }
    \label{fig1}
\end{figure}

The Hamiltonian of $t-t^{\prime}-U$ Hubbard model on a square lattice is given by
\begin{equation}
H =-\sum_{\substack{i, l, \sigma}} \left( t_{l, \sigma} c_{i,\sigma}^\dag c_{i+l,\sigma} + h.c. \right)
+ U \sum_i n_{i,\uparrow} n_{i,\downarrow} 
\label{eq:hamiltonian}
\end{equation}
where $c_{i,\sigma}^\dag$ ($c_{i,\sigma}^{\,}$) is electron creation (annihilation) operators with spin $\sigma$ = $\uparrow, \downarrow$, and $n_{i,\sigma}=c_{i,\sigma}^\dag c_{i,\sigma}$ is the electron number operator. The electron hopping amplitude $t_{l, \sigma}$, where $l = \pm \hat{x}, \pm \hat{y}$ represents the NN sites of a given site $i$, and $l = \pm \hat{x} \pm \hat{y}$ represents the NNN sites of a given site $i$. $U < 0$ is the on-site attractive interaction. For defining the spin-dependent anisotropic hopping amplitudes, we define the variable $\alpha$ and $\alpha$', where $t_{\hat{y}\downarrow} = t_{\hat{x}\uparrow} = t$, $t_{\hat{x}\downarrow} = t_{\hat{y}\uparrow} = \alpha t$ (we set the NN hopping $t$ = 1 as the energy unit) and $t_{\hat{x}-\hat{y}\downarrow} = t_{\hat{x}+\hat{y}\uparrow} = t^{\prime}$, $t_{\hat{x}+\hat{y}\downarrow} = t_{\hat{x}-\hat{y}\uparrow} = \alpha' t^{\prime}$, leading to an unpolarized system with balanced spin populations $\langle n_{i,\uparrow} \rangle = \langle n_{i,\downarrow} \rangle = n/2$.
We give a schematic diagram with anisotropic NN and NNN hopping on 2D square lattice (see Fig.\ \ref{fig1}a), without loss of generality we defined the anisotropy parameter $\alpha, \alpha'  \in [0,1]$. Such a Hubbard model with unequal hopping amplitudes can be readily implemented in an optical lattice by loading mixtures of ultracold fermionic atoms with different masses. The correlation functions of CPQMC are discussed in the Supplemental Material \cite{Cao2024t-supp}.

CPBM is an exotic non-superfluid paired state of fermions, the uncondensed Cooper pairs would form a collective state with low-energy gapless excitations along a Bose surface \cite{Feiguin2009-nu,Feiguin2011-ib}. In our recent work \cite{Cao2024-qd}, we explore a diverse phase diagram as a function of electron density $n$ and anisotropy $\alpha$ in $t-U$ model, and reveal the emergence of a CPBM phase in a highly anisotropic regime with wide range of filling. 
These experiences are also applied to the $t-t^{\prime}-U$ model, so we now focus on a case with electron density $n \sim 0.65$ at $U=-3$, and explore the influence of NNN hopping $t^{\prime}$ on the ground state properties of the Bose metal phase. The CPBM phase in this parameter range is relatively stable on different lattice sizes. In this paper, we assume $\alpha=\alpha'$ for simplicity.

As we demonstrate in Fig.\ \ref{fig1}(b), we observe a diverse phase diagram for different value of NNN hopping $t^{\prime}$ and anisotropy $\alpha$ with electron filling $n \sim 0.65$ at zero temperature. Overall, the exotic CPBM phase still exists in a strong anisotropy, emerging when a spin-dependent anisotropy suppresses the ordinary $s$-wave superfluid ($s$-SF). Fig.\ \ref{fig1}(c)-(e) shows the Bose surface in momentum space representing different values of $t^{\prime}=0.2,\ 0.5,\ 0.8$ at extreme anisotropy $\alpha=0.05$. IDW is an incommensurate density wave at extremely strong anisotropy range, and it disappears when $t^{\prime}>0.4$. We can see that the Bose surface still exists, and the appearance of $t^{\prime}$ does not significantly change the main properties of the CPBM phase, but it changed the shape and size of the Bose surface, which is mainly related to the Fermi surface distortion caused by spin anisotropy. However, the region of CPBM phase is affected by $t^{\prime}$, we can see from the phase diagram that as $t^{\prime}$ increases, the area where CPBM exists presents a dome-like shape, the maximal area of CPBM phase occurs at around the optimal strength of $t^{\prime}\sim 0.2$, and then decreases in both small and large $t^{\prime}$ regions. This indicates that the presence of the appropriate value of $t^{\prime}$ is beneficial for Bose metal phase.  Further details on determining the various phase regions in the phase diagram from CPQMC data are presented
in the Supplemental Material \cite{Cao2024t-supp}.

\begin{figure}[b!]
    \centering
    \includegraphics[width=1\linewidth]{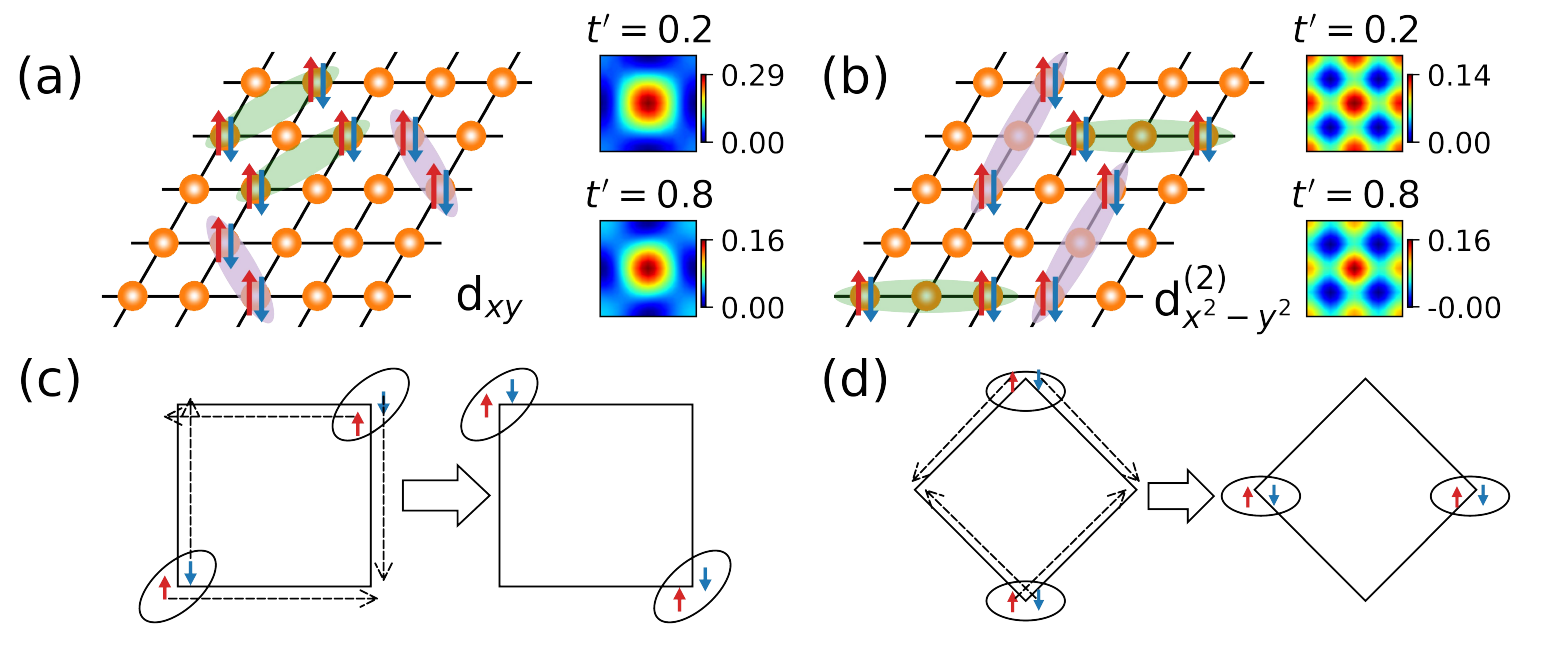}
    \caption{(Color online) $d$-wave boson correlation. (a) and (b) (left panel) The schematic of the two-boson (on site pairing) $d_{xy}$-wave and $d^{(2)}_{x^2-y^2}$-wave pairing in square lattice. (right panel) Two-boson correlator $P_{d_{xy}-boson}({\bf k})$, $P_{d^{(2)}_{x^2-y^2}-boson}({\bf k})$ for $t^{\prime} = 0.2$ and $0.8$ at $\alpha =0.05$ with $n \sim0.65$ on 16 $\times$ 16 lattice, respectively. (c) Processes that would give origin to an effective ring-exchange $K_t$ of pairs induced by $t$ anisotropy. (left) Fermions with opposite spin prefer to move along perpendicular directions. (d) Processes that would give origin to an effective ring-exchange $K_{t^{\prime}}$ of pairs induced by $t^{\prime}$ anisotropy. (left) Fermions with opposite spin prefer to move along diagonal directions perpendicularly. }
    \label{fig2}
\end{figure}

\begin{figure}[b!]
    \centering
    \includegraphics[width=1\linewidth]{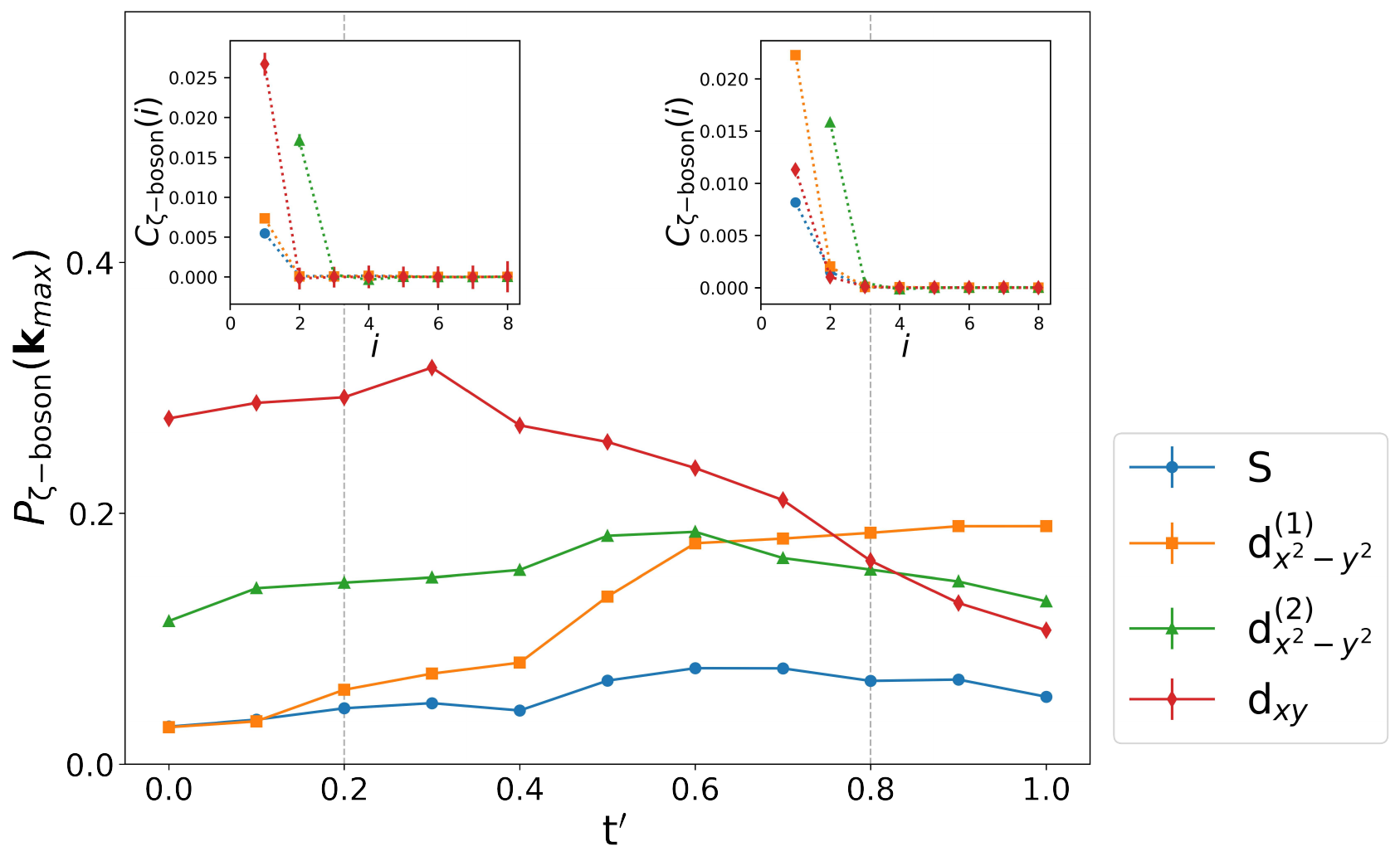}
    \caption{(Color online) Competition between $d_{x^2-y^2}$-wave and $d_{xy}$-wave boson correlation. The strength of four pairing modes of bosons paired correlation at $\alpha =0.05$ with $n \sim0.65$ varies with different $t^{\prime}$. The inset shows boson pairing correlations in the real space for $t^{\prime}= 0.2$ and $0.8$, marked by the black dashed lines. }
    \label{fig3}
\end{figure}

One of the most important prospect of the Bose metal phase is its $d$-wave correlation between bosons (on-site pairing) \cite{Feiguin2011-ib,Cao2024-qd}. We defined the two-boson correlator $P_{\mathrm \zeta-boson}({\bf k})$ to study the possibility of a phase of Cooper pairs. Here $\zeta = s,\ d_{xy},\ d^{(1)}_{x^2-y^2},\ d^{(2)}_{x^2-y^2} $.
Among them, the $d_{x^2-y^2}$-wave is divided into two types, namely the NN and the third NN $d_{x^2-y^2}$-wave, which we use $d^{(1)}_{x^2-y^2}$-wave and $d^{(2)}_{x^2-y^2}$-wave to represent. 
Our previous results \cite{Cao2024-qd} suggest that the correlation between Cooper pairs is predominantly $d_{xy}$-wave in the CPBM regime when there is no $t^{\prime}$ in 2D system.

\begin{figure*}
    \centering
    \includegraphics[width=1\linewidth]{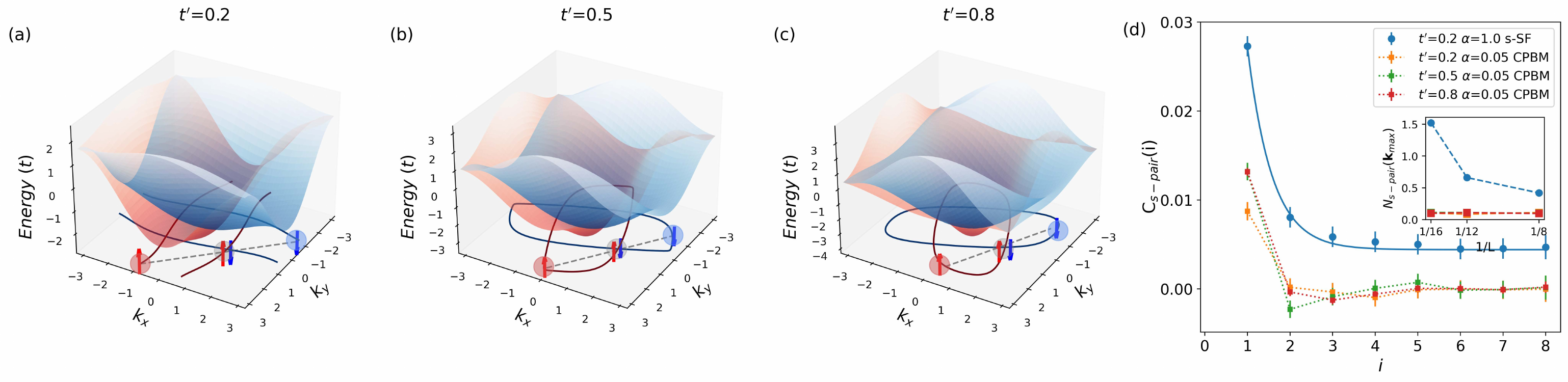}
    \caption{(Color online) Characteristic of CPBM phase. (a)-(c) Non-interacting energy spectrum of the spin-dependent Fermi surface with anisotropy $\alpha =0.05$ for $t^{\prime} = 0.2,\ 0.5,\ 0.8$ at filling $n\sim 0.65$. The Fermi surface is projected as red~(blue) lines for the up-spin~(down-spin). We provide an illustration of pairing with nonzero momentum. (d) The $s$-wave pairing correlation function in the CPBM phase and the $s$-SF phase in real space. The solid blue line is an exponential fitting curve. The inset is lattice size effect in CPBM phase and $s$-SF phase.}
    \label{fig4}
\end{figure*}

In left panel of Fig.\ \ref{fig2}(a) and Fig.\ \ref{fig2}(b), we give the schematic of the two-boson $d_{xy}$-wave pairing correlations on the diagonals in square lattice and $d^{(2)}_{x^2-y^2}$-wave pairing correlations on the third NN sites. The $d$-wave here specifically refers to $d_{xy}$-orbital or $d_{x^2-y^2}$-orbital symmetry, which exhibits a propensity to introduce $d$-wave correlations into the system and qualitatively alter the sign structure of the electronic ground state. We show the CPQMC simulation results of two-boson correlator $P_{\mathrm \zeta-boson}({\bf k})$ in momentum space for $\alpha =0.05$ at different $t^{\prime}=0.2$ and $t^{\prime}=0.8$, $\zeta = \ d_{xy},\ d^{(2)}_{x^2-y^2}$, respectively.
In the highly anisotropic limit, the process depicted in Fig.\ \ref{fig2}(c) and (d) dominates, corresponding to $d_{xy}$-wave and $d^{(2)}_{x^2-y^2}$-wave boson correlations induced by $t$ and $t^{\prime}$, respectively.

By model mapping, we provided the connection between our model of fermions with spin-dependent anisotropic Fermi surface to the bosonic $J-K$ model. When $\left| U \right| \gg t$, all of the fermions are considered to be tightly bound into on-site Cooper pairs. We can derive an effective boson Hamiltonian by considering a perturbation expansion in powers of $t/\left| U \right| $ \cite{Motrunich2007-rb,Feiguin2009-nu,Cao2024-qd,Feiguin2011-ib}.  For $t-t^{\prime}-U$ model, the effective Hamiltonian can be written as: $H_b = -J\sum_{i, j}b_{i}^\dag b_{j} + K_t\sum_{ring}b_{1}^\dag b_{2}b_{3}^\dag b_{4} + K_{t^{\prime}}\sum^{\prime}_{ring}b_{1}^\dag b_{2}b_{3}^\dag b_{4} + h.c.$. Here, we consider the boson ring-exchange term $K_t$ and $K_{t^{\prime}}$ as depicted in Fig.\ \ref{fig2}(c) and (d), which involves two bosons on opposite corners of an elementary square plaquette and rhomboid plaquette rotating by $\pm 90$ degrees. Here, $b_{i}^\dag = c_{i \uparrow}^\dag c_{i \downarrow}^\dag$, $K_t$ corresponds to $i = 1, 2, 3, 4$ labeling sites taken clockwise around a square plaquette, $K_{t^{\prime}}$ corresponds to $i = 1, 2, 3, 4$ labeling sites taken clockwise around a rhomboid plaquette. Remarkably, in the extreme anisotropic limit, the ring terms $K_t$ and $K_{t^{\prime}}$ which hops pairs of bosons is nonzero, they can even be comparable to the spin exchange coupling term $J$. At the same time as the $K_{t^{\prime}}$ ring exchange term induced by $t^{\prime}$ anisotropy, the boson in the ring exchange may also have a boson correlation with another NN boson. In addition, we also provide the physical mechanism for enhancing $d^{(1)}_{x^2-y^2}$-wave by $K_{t^{\prime}}$ in Supplemental Material \cite{Cao2024t-supp}. Particularly, with increasing $t^{\prime}$, the coupling strength $K_{t^{\prime}}$ becomes increasingly more important in the system, leading to the $d^{(1)}_{x^2-y^2}$-wave and $d^{(2)}_{x^2-y^2}$-wave all becoming more important.

We demonstrated changes in the strength of four pairing modes of boson paired correlation by  fixing $\alpha =0.05$ at $n\sim 0.65$ and changing the NNN hopping $t^{\prime}$ (Fig.\ \ref{fig3}). Overall, the primary competition arises between $d_{xy}$-wave and two types of $d_{x^2-y^2}$-wave, while $s$-wave holds relatively less significance with a small value and the three types of $d$-waves have relatively high values in CPBM region. As $t^{\prime}$ increases, the intensity of the $d^{(1)}_{x^2-y^2}$-wave boson correlation increases and the $d^{(2)}_{x^2-y^2}$-wave does not change much, the proportion of the $d_{xy}$-wave boson correlation decreases, and ultimately the $d_{x^2-y^2}$-wave dominates when $t^{\prime}>0.7$. 
In the inset of Fig.\ \ref{fig3}, we compare the correlations between four pairing modes of boson pairs in real space for $t^{\prime} = 0.2$ and $0.8$ (the parameters consistent with right panel of Fig.\ \ref{fig2}(a) and Fig.\ \ref{fig2}(b)). They all decay rapidly to zero, indicating that the boson correlation is short-range. The strength of the $d_{xy}$-wave boson correlation is the {strongest when $t^{\prime}=0.2$, while two types of $d_{x^2-y^2}$-wave dominate when $t^{\prime}=0.8$. Overall, we suggest its predominant boson pairing mode is always $d$-wave character in Bose metal phase. Particularly, our CPQMC results indicate that the system tends toward $d_{x^2-y^2}$-wave correlation between Cooper pairs when $t^{\prime}$ is large ($t^{\prime}>0.7$), and $t^{\prime}$ being a key tuning parameter to adjust the intensity of boson correlation in the boson metallic phase. This exotic phase of boson-paired states can be termed the $d$-wave Bose Metal \cite{Sheng2008-jd,Feiguin2011-ib,Jiang2013-io,Cao2024-qd}.

We show the non-interacting energy dispersion with anisotropy $\alpha =0.05$ for $t^{\prime} = 0.2,\ 0.5,\ 0.8$ in Fig.\ \ref{fig4}(a)-(c). The solid red and blue lines below indicate the anisotropic Fermi surface of spin-up and spin-down for a projection at $n \sim 0.65$. Due to the severe mismatch Fermi surface, fermions with different spin near the Fermi surface favor forming pairs at finite momentum. The previous mean-field study restricts the pairing between patches of fermions with opposite spin located along a specific direction~\cite{Motrunich2007-rb}. Here our quantum Monte Carlo result considers the influence of all possible pairing channel with the restriction. We provide an illustration of nonzero momentum pairing in Fig.\ \ref{fig4}(a)-(c), respectively. Pairs with different finite momentum are formed by the spin-up and spin-down fermions at different parts of anistropic Fermi surface, constituting a continuous Bose surface with singular pair distribution function, as shown in Fig.\ \ref{fig1}(c)-(e), the maximum weight of pairing is relatively evenly distributed on the Bose surface. The shape and sharpness of the Bose surface are regulated by both the filling $n$ and hopping anisotropy $\alpha$ and the NNN hopping $t^{\prime}$.

The presence of anisotropy $\alpha$ induces a mismatch between the spin-up and spin-down Fermi surfaces, while $t^{\prime}$ distorts the shape of the Fermi surface. As $t^{\prime}$ increases, this distortion intensifies. When the distortion is significant, the rotation of the anisotropic Fermi surface from the $t^{\prime} = 0$ limit becomes pronounced. At the $t^{\prime} = t$ limit, the elliptic Fermi surfaces almost orient along the $k_x \pm k_y$ direction. Consequently, the resulting Bose surface in the CPBM phase also rotates accordingly, as illustrated. This leads to closed Fermi surfaces forming at small to medium filling. Previous studies have pointed out that the openness of the Fermi surfaces is crucial in classifying the Bose metal, resulting in $d$-wave Local Bose Liquid (DLBL) with an open Fermi surface and $d$-wave Bose Liquid (DBL) with a closed Fermi surface~\cite{Motrunich2007-rb}. The analogous CPBM phase to DBL can be realized by introducing a sufficiently large $t^{\prime}$, whereas previous studies of CPBM mostly correspond to DLBL due to the open nature of the Fermi surfaces.

Fig.\ \ref{fig4}(d) compares the CPQMC simulation results of the pair correlation function for CPBM phase and $s$-SF phase in real space. In particularly, the correlation in real space decay to zero quickly and fluctuate around zero with increasing distance in CPBM phase, showing that correlation is short-range, and there is no significant difference as $t^{\prime}$ changes. On the contrary, the correlation exhibits an exponential decay, and converges to a steady finite value at long distance in $s$-SF phase, this is the hallmark of the long-range superconducting correlation among Cooper pairs. 
Inset of Fig.\ \ref{fig4}(d) show the lattice size effect in different phases, we can see that in CPBM region when $\alpha = 0.05$, the value of $N_{\mathrm s-pair}({\bf k}_{\rm max})$ has always been stable, which represents that the CPBM phase will always exist in large or even infinite lattice size. When $\alpha = 1.00$, the system is $s$-SF phase, the value of $N_{\mathrm s-pair}({\bf k}_{\rm max})$ is continuously increasing, so $s$-SF is significantly diverge at weak anisotropic region.

In recent years, charge order or charge density wave (CDW)—a static periodic modulation of charge density and lattice positions driven by the Fermi surface, has been shown to be a universal property of cuprates \cite{comin2016resonant} and other unconventional superconductors \cite{Teng2022-md}.
IDW along the lattice diagonals exists at strong anisotropy, displaying singular features condenses at nonzero momentum point $\bf Q$ = ($2k_F$, $2k_F$), gradually diffusing to point $\bf Q$ = ($\pi$, $\pi$) convert to CDW as $n$ approached to half filling \cite{Cao2024-qd}.
We also explore the influence of $t^{\prime}$ on density wave and associated periodicity. We demonstrate that IDW disappears when $t^{\prime}>0.4$ in phase diagram, suggesting that too large $t^{\prime}$ is not conducive to the formation of IDW. We have discussed the influence of $t^{\prime}$ on density wave and associated periodicity in details in the Supplemental Material \cite{Cao2024t-supp}.

To conclude, we utilize the CPQMC algorithm to study the effect of the NNN hopping $t^{\prime}$ on the CPBM phase and its boson correlation in 2D lattice. In the highly anisotropic regime ($\alpha < 0.40$) at filling n$\sim$0.65 for various  $t^{\prime}$ values, we observe the presence of Bose surface in Cooper-pair distribution function, which is compelling evidence of the CPBM phase. Furthermore, the CPBM region in phase diagram exhibits a dome-like shape, with the maximal CPBM phase region occurring at around the optimal strength of  $t^{\prime}\sim 0.2$, and then decreases in both small and large $t^{\prime}$ regions. Subsequently, we explored the existence of boson correlation, suggesting that the boson correlation between Cooper pairs is predominantly $d_{x^2-y^2}$-wave in the CPBM regime with large $t^{\prime}$ ($t^{\prime}>0.7$).
We also argued that the necessary condition to form CPBM phase is the anisotropy of the Fermi surface caused by strong spin anisotropy.

Recently, spin-dependent anisotropic Fermi surfaces have garnered significant attention in condensed matter physics, particularly in the context of altermagnetism~\cite{Smejkal2022-wp,Smejkal2022-po}.
A recent study theoretically proposed how a $d$-wave altermagnetic phase can be realized with ultracold fermionic atoms in optical lattices, which in a altermagnetic Hubbard model with anisotropic NNN hopping $t^{\prime}$ \cite{Das2023-ng}. The $d_{x^2-y^2}$-wave Bose metal phase can be realized in optical lattice experiments by tuning the filling and hopping anisotropy of effective spin interactions with light.
Our results demonstrate that the properties of Bose metal phase is strongly affected by modifications of the NNN hopping $t^{\prime}$ in Hubbard model. This provides valuable guidance for ultracold atomic gases in optical lattices, which can be microscopically engineered and measured to investigate these exotic phases within specific parameter ranges.

This work is supported by the National Natural Science Foundation of China~(Grant No. 12204130), Shenzhen Start-Up Research Funds~(Grant No. HA11409065), HITSZ Start-Up Funds~(Grant No. X2022000), Shenzhen Key Laboratory of Advanced Functional CarbonMaterials Research and Comprehensive Application (Grant No. ZDSYS20220527171407017). T.Y. acknowledges supports from Natural Science Foundation of Heilongjiang Province~(No.~YQ2023A004).

\bibliography{ref}
\end{document}

\title{Supplemental Material for "$d_{x^2-y^2}$-wave Bose Metal induced by the next-nearest-neighbor hopping $t^{\prime}$"}

\author{Zhangkai Cao}
\thanks{These authors contributed equally.}
\affiliation{School of Science, Harbin Institute of Technology, Shenzhen, 518055, China}

\author{Jianyu Li}
\thanks{These authors contributed equally.}
\affiliation{School of Science, Harbin Institute of Technology, Shenzhen, 518055, China}
\affiliation{Shenzhen Key Laboratory of Advanced Functional Carbon Materials Research and Comprehensive Application, Shenzhen 518055, China.}
 
\author{Jiahao Su}
\affiliation{School of Science, Harbin Institute of Technology, Shenzhen, 518055, China}
\affiliation{Shenzhen Key Laboratory of Advanced Functional Carbon Materials Research and Comprehensive Application, Shenzhen 518055, China.}

\author{Tao Ying}
\affiliation{School of Physics, Harbin Institute of Technology, Harbin 150001, China}
 
\author{Ho-Kin Tang}
\email{denghaojian@hit.edu.cn}
\affiliation{School of Science, Harbin Institute of Technology, Shenzhen, 518055, China}
\affiliation{Shenzhen Key Laboratory of Advanced Functional Carbon Materials Research and Comprehensive Application, Shenzhen 518055, China.}

\date{\today} 
\maketitle 

In this Supplemental Material, we present additional definitions and computational results. In Sec.~S1, we have defined various correlation functions. In Sec.~S2, we give two possible mechanism leading to the enhancement of the two-boson $d^{(1)}_{x^2-y^2}$-wave pairing correlation in the nearest neighbor. In Sec.~S3, we show the transition of $s$-SF phase to CPBM phase induced by anisotropy in different $t^{\prime}$. In Sec.~S4, we discuss the determination of the CPBM phase. In Sec.~S5, we discuss the boson correlation of various symmetry. In Sec.~S6, we investigated the competition between $d$-wave boson correlation when $t<t^{\prime}$. In Sec.~S7, we discuss the incommensurate density wave in the small $t^{\prime}$ limit. In Sec.~S8, we give a brief introduction to Constraint path quantum Monte Carlo.

\section{Correlation function}

We defined the on-site $s$-wave pair momentum distribution function
\begin{equation}
N_{\mathrm s-pair}({\bf k}) = (1/N)\sum_{i,j} \mbox{exp}[i{\bf k}({\bf r}_i-{\bf r}_j)]\, \langle {\Delta}_{s}^{\dagger}(i) {\Delta}_{s}(j) \rangle,
\label{nspdtkpair}
\end{equation}
Here, $N$ is the number of sites, and $\Delta_s^\dagger(i)=c^\dagger_{i\uparrow}c^\dagger_{i\downarrow}$. 

To study the density correlations, we defined the charge structure factor in particle-hole channel, 
\begin{equation}
N_{\mathrm c}({\bf k}) = (1/N)\sum_{i,j} \mbox{exp}[i{\bf k}({\bf r}_i-{\bf r}_j)]\, \langle {n}_i {n}_j \rangle,
\label{nskpair}
\end{equation}
Here, the density number operator is defined as
${n}_i = \sum_{\sigma} c^\dagger_{i \sigma} c_{i \sigma}$.

The correlation function of on site $s$-wave pairing mode in real space are defined as $C_{\mathrm s} = \langle {\Delta}_{s}^{\dagger}(i) {\Delta}_{s}(j) \rangle$.
The correlation function of charge density wave in real space are defined as $C_{\rm CDW} = \langle {n}_i {n}_j \rangle$.

We also defined the two-boson (on site pairing) correlator,
\begin{equation}
P_{\mathrm \zeta-boson}({\bf k}) = (1/N)\sum_{i,j} \mbox{exp}[i{\bf k}({\bf r}_i-{\bf r}_j)]\, \langle {b}^{\dagger}(i) {b}^{\dagger}(i') {b}(j) {b}(j') \rangle,
\label{Pnspdtkpair}
\end{equation}
where $\zeta = s,\ d_{xy},\ d^{(1)}_{x^2-y^2},\ d^{(2)}_{x^2-y^2} $. When $\zeta = s,\ d^{(1)}_{x^2-y^2}$, $i'$ or $j'$ site is the NN sites of $i$ or $j$ site, and when $\zeta = d_{xy}$, $i'$ or $j'$ site is the NNN sites of $i$ or $j$ site, and when $\zeta = d^{(2)}_{x^2-y^2}$, $i'$ or $j'$ site is the third NN sites of $i$ or $j$ site. 
Among them, we focus more attention on the $d_{xy}$-wave and two types of $d_{x^2-y^2}$-wave boson correlation. The boson pairing correlations in real space are defined as $C_{\mathrm \zeta-boson} = \langle {b}^{\dagger}(i) {b}^{\dagger}(i') {b}(j) {b}(j') \rangle$, where $\zeta = s,\ d_{xy},\ d^{(1)}_{x^2-y^2},\ d^{(2)}_{x^2-y^2} $.

\section{Possible mechanism for the enhancement of $d^{(1)}_{x^2-y^2}$-wave boson correlation}

We have already shown the physical mechanism of $d_{xy}$-wave and $d^{(2)}_{x^2-y^2}$-wave in Fig. 2 of the main text. But in detailed measurements, we found that with $t^{\prime}$ increasing, pairing of $d^{(1)}_{x^2-y^2}$-wave shows the most significant growth. We provide its possible mechanism below.

\begin{figure*}[htb!]
    \centering
    \includegraphics[width=0.8\linewidth]{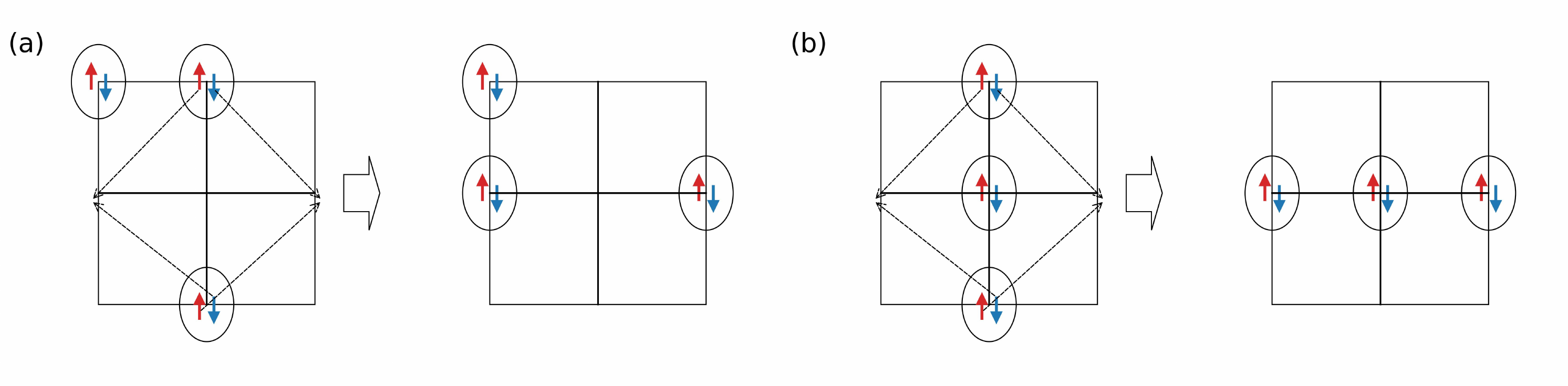}
    \caption{(Color online) (a) and (b) Two possible processes contributing to the enhancement of $d^{(1)}_{x^2-y^2}$-wave pairing correlations on the NN sites, which are closely related to the effective ring-exchange $K_{t^{\prime}}$ of pairs induced by $t^{\prime}$ anisotropy. (left) Fermions with opposite spin prefer to move along diagonal directions perpendicularly, which not only lead to $d^{(2)}_{x^2-y^2}$-wave pairing with the third NN bosons, but also results in the effective $d^{(1)}_{x^2-y^2}$-wave pairing with the NN bosons.}
    \label{figS1}
\end{figure*}

When $\left| U \right|$ is relatively large, the effective Hamiltonian can be written as: $H_b = -J\sum_{i, j}b_{i}^\dag b_{j} + K_t\sum_{ring}b_{1}^\dag b_{2}b_{3}^\dag b_{4} + K_{t^{\prime}}\sum^{\prime}_{ring}b_{1}^\dag b_{2}b_{3}^\dag b_{4} + h.c.$. 
In the extreme anisotropic limit, the ring terms $K_t$ and $K_{t^{\prime}}$ can be comparable to the spin exchange coupling term $J$. The impact of these two ring exchange terms is significant. As noted, the $d_{xy}$-wave and $d^{(2)}_{x^2-y^2}$-wave are caused by the ring exchange terms $K_t$ and $K_{t^{\prime}}$. However, there is also an important contribution from nearest-neighbor $d^{(1)}_{x^2-y^2}$-wave pairing, which is closely attributed to the $K_{t^{\prime}}$ term. In Fig.~\ref{figS1}, we present two possible enhancement mechanisms.

As the $K_{t^{\prime}}$ ring exchange term is induced by $t^{\prime}$ anisotropy, the boson involved in the ring exchange process is also correlated with another nearest-neighbor boson. When the bosons in the ring exchange term undergo exchange, in addition to the $d^{(2)}_{x^2-y^2}$-wave boson correlation of the two bosons in the ring exchange term, the boson correlation between a boson in the ring exchange and another nearest-neighbor boson also contributes to the $d^{(1)}_{x^2-y^2}$-wave channel. Thus, these processes contribute to the increase of the $d^{(1)}_{x^2-y^2}$-wave pairing correlations with increasing $t^{\prime}$.

\section{Transition from $s$-SF phase to CPBM phase induced by anisotropy}

\begin{figure*}[hb!]
    \centering
    \includegraphics[width=1\linewidth]{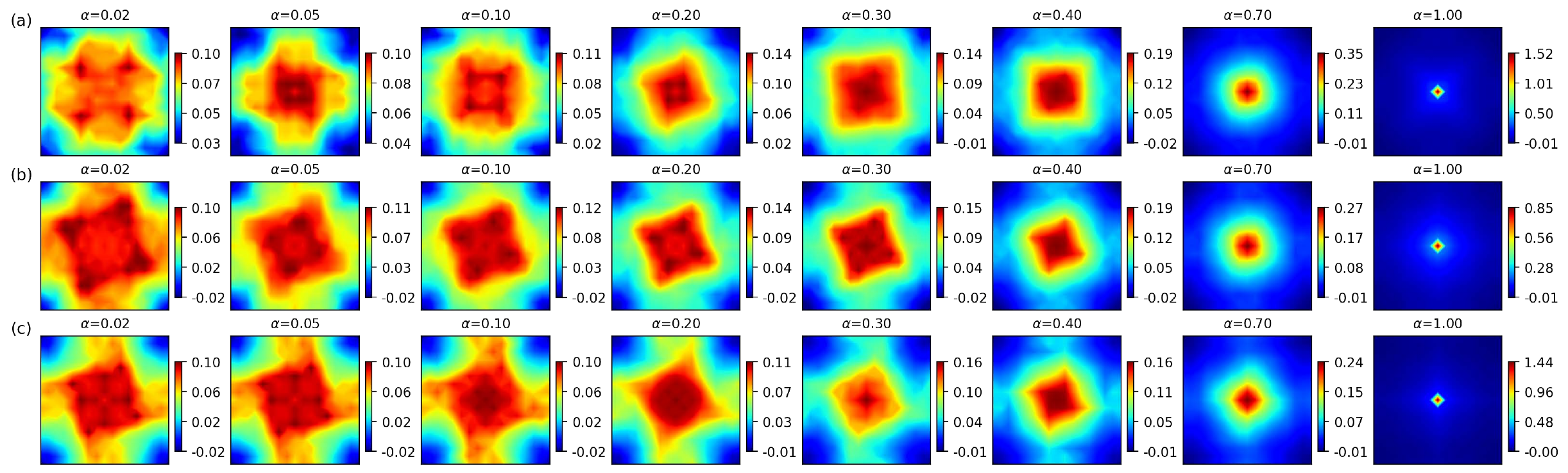}
    \caption{(Color online) Transition of $s$-SF to CPBM induced by anisotropy.  CPQMC simulation result of $N_{\mathrm s-pair}({\bf k})$ from extremely anisotropy $\alpha \rightarrow 0$ to isotropic $\alpha = 1$ at $n \sim 0.65$ for (a) $t^{\prime}=0.2$, (b) $t^{\prime}=0.5$, (c) $t^{\prime}=0.8$ on 16 $\times$ 16 lattice. }
    \label{figS2}
\end{figure*}

In this section, we demonstrate the transition of $s$-SF phase to CPBM phase induced by anisotropy $\alpha$ in $t-t^{\prime}-U$ model. We give the CPQMC simulation result of $N_{\mathrm s-pair}({\bf k})$ from extremely anisotropy $\alpha \rightarrow 0$ to isotropic $\alpha = 1$ in Fig.\ \ref{figS2} for $t^{\prime}$= 0.2, 0.5, 0.8. We see that nonzero momentum Bose surface only appear in strong anisotropy, and as anisotropy increases ($\alpha \downarrow$), a more pronounced and larger complete Bose surface is formed. When the anisotropy is weak, the system exhibits $s$-SF condensation at $\bf Q$ = (0, 0). In the moderate anisotropic region, there are also weakened $s$-SF, which is slightly suppressed at moderately anisotropic region. As $t^{\prime}$ increases, the shape of Bose surface in CPBM phase will rotate, and the size of Bose surface will also increase.

On the whole, as the anisotropy increases, the difference of the Fermi surface by the two spin electrons also increases, leading to a tendency towards pairing at larger ${\bf k}_{pair}$ in the Brillouin zone, thus forming the exotic CPBM. The transition of $s$-SF to CPBM induced by anisotropy is very natural, and from weak $s$-SF to CPBM phase, the system undergoes no symmetry breaking. In addition, the size and shape of the Boson surface is closely related to the anisotropy $\alpha$ and the NNN hopping $t^{\prime}$. 

\section{Determination of the CPBM phase}

This section mainly introduces how to determine CPBM phase regions in the phase diagram from the CPQMC data.

\begin{figure*}[htb!]
    \centering
    \includegraphics[width=0.8\linewidth]{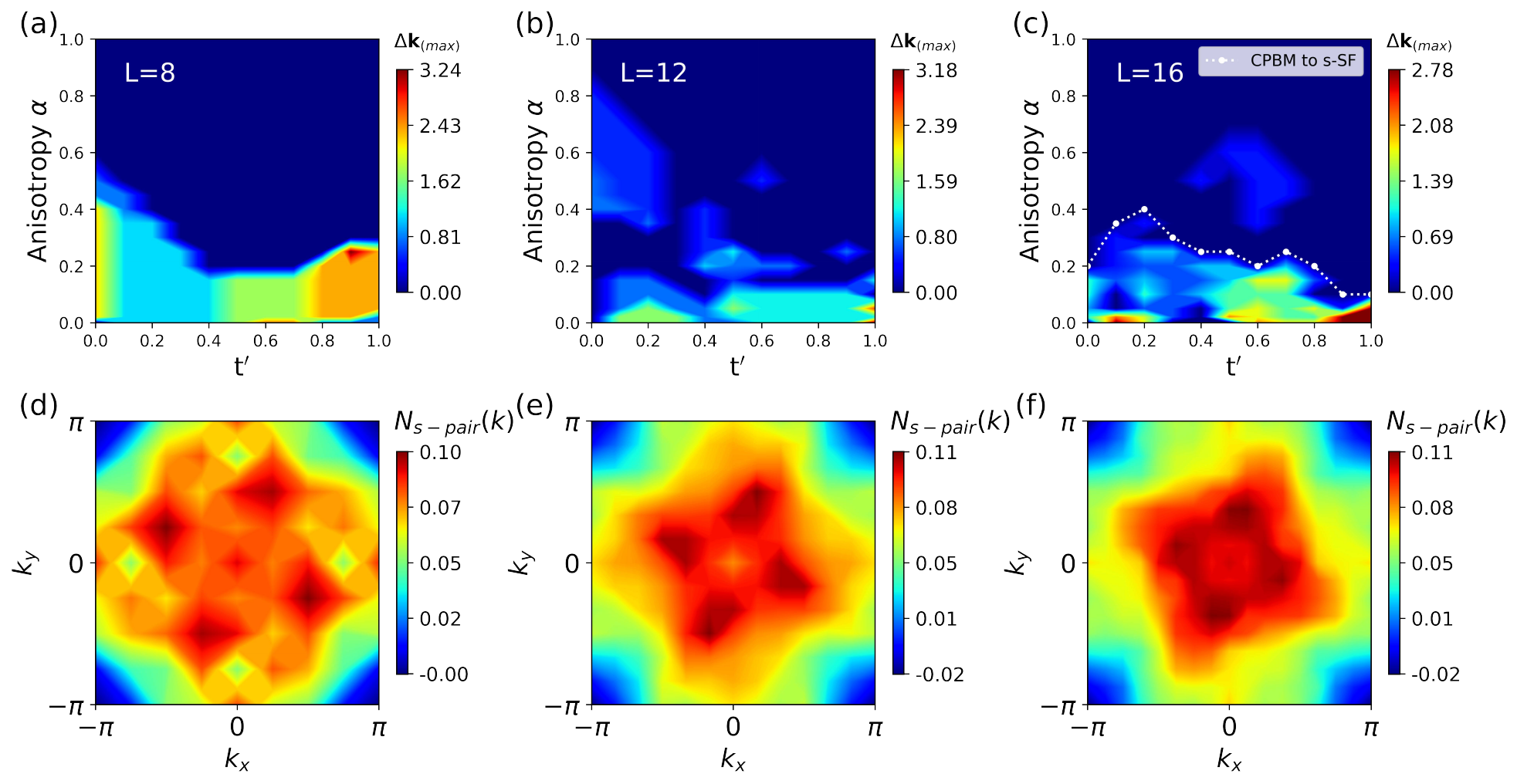}
    \caption{(Color online) Determination of the CPBM. (a)-(c) CPQMC simulation result for determine the region of CPBM at $U= -3$ on 8 $\times$ 8, 12 $\times$ 12, 16 $\times$ 16 lattice, where $\Delta {\bf k}_{\rm max}=\sqrt{{\bf k}_{\rm max}^2(x)+{\bf k}_{\rm max}^2(y)}$, ${\bf k}_{\rm max}(x)$, ${\bf k}_{\rm max}(y)$ are the coordinates of the maximum value in the pair momentum distribution function. The blue areas are $\Delta {\bf k}_{\rm max} = 0$, representing the $s$-wave pair momentum distribution function condensed at $\bf Q$ = (0, 0), other regions are $\Delta {\bf k}_{\rm max} \neq 0$. If $\Delta {\bf k}_{\rm max} \neq 0$, it indicates that there is a nonzero momentum peak in the pair momentum distribution function, and then combined with image display to form a Bose surface, so we believe that there exist a CPBM phase under this parameter. (d)-(f) CPQMC simulation result of $s$-wave pair momentum distribution function $N_{\mathrm s-pair}({\bf k})$ for $\alpha =0.05$ at $n$ $\sim$ 0.65 with $t^{\prime}= 0.5$ in $U= -3$ on 8 $\times$ 8, 12 $\times$ 12, 16 $\times$ 16 lattice, showing obviously existence of nonzero momentum Bose surface, which is the signal for determine the exotic CPBM phase.  }
    \label{figS3}
\end{figure*}

We have shown the determination of the CPBM region in Fig.\ \ref{figS3}(a)-(c), on 8 $\times$ 8, 12 $\times$ 12, 16 $\times$ 16 lattice, with non-zero peak in $N_{s-pair}$. We can see at small lattice size, the distribution of CPBM region is relatively unstable and fluctuates, but as the lattice size increases, the Bose surface gradually stabilizes. Some weak non-zero peak signals in weak anisotropy are believed to be caused by lattice size effect.
Because of constraints in computational resources availability, our calculations were limited to a maximum lattice size of 16 $\times$ 16. In Fig.\ \ref{figS3}(c), we have drawn the boundary from CPBM phase to s-SF phase, based on the results of 16 $\times$ 16 lattice, as shown in phase diagram of main text. From the trend of change, we predict that the CPBM region under infinite lattice points N $\rightarrow\infty$ will be connected as a whole, reaching its optimal strength of $t^{\prime}\sim 0.2$ and decreasing in both smaller and larger $t^{\prime}$ regions. Meanwhile, we draw the Bose surface of features in CPBM phase at different lattice size in Fig.\ \ref{figS3}(d)-(f), selecting on different lattice for $\alpha =0.05$ at $n$ $\sim$ 0.65 with $t^{\prime}= 0.5$. As the lattice size increases, the nonzero momentum Bose surface persists and its shape becomes conspicuous. Fig.\ \ref{figS2} shows the transition of s-SF phase to CPBM phase caused by anisotropy $\alpha$ in different $t^{\prime}$.

\section{Boson correlation}

\begin{figure*}[ht!]
    \centering
    \includegraphics[width=1\linewidth]{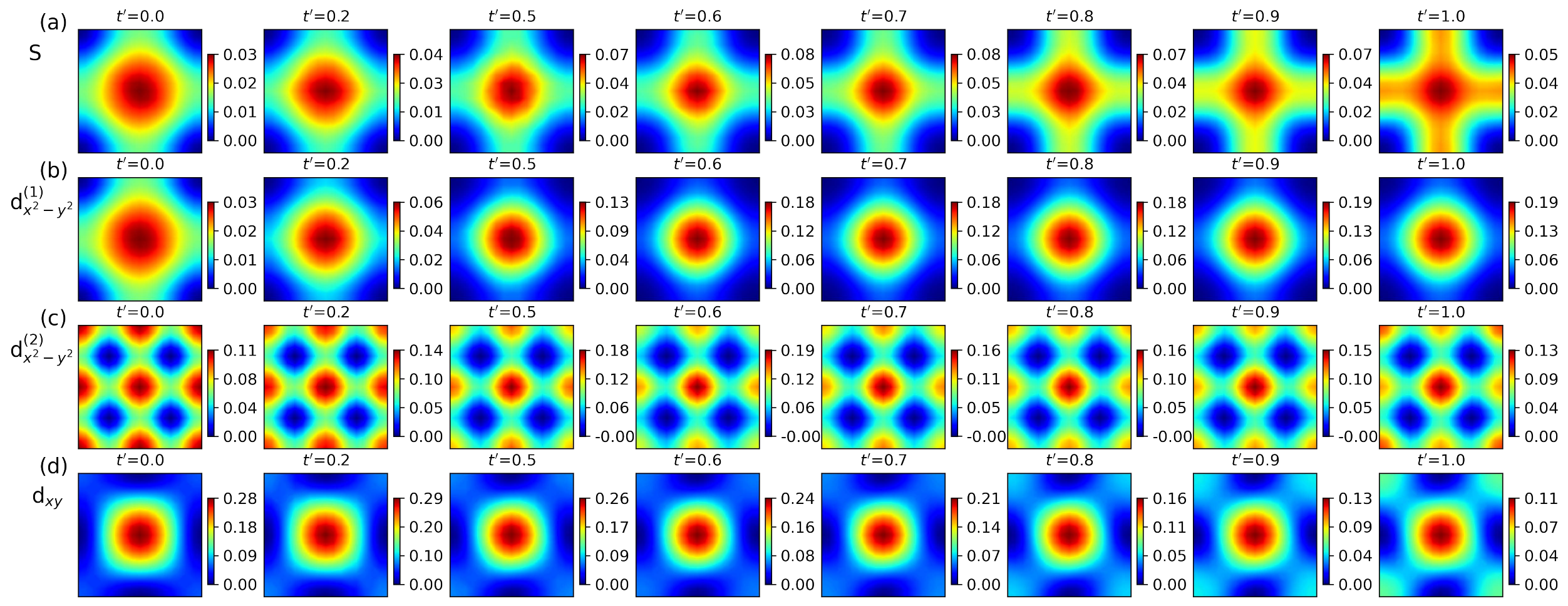}
    \caption{(Color online) Boson correlation for different $t^{\prime}$. (a)-(d) CPQMC simulation results of two-boson correlator $P_{\mathrm \zeta-boson}({\bf k})$ in momentum space for $t^{\prime}=$ 0.0, 0.2, 0.5, 0.6, 0.7, 0.8, 0.9, 1.0 with anisotropy $\alpha=0.05$ at $n\sim 0.65$, $\zeta = s,\  d^{(1)}_{x^2-y^2},\ d^{(2)}_{x^2-y^2},\ d_{xy}$, respectively.   }
    \label{figS4}
\end{figure*}

In CPBM phase, the electrons near the Fermi surface are pairing on-site in the system. Therefore, it is important to explore the correlation between bosons, especially when the system does not exhibit superfluid condensation in CPBM phase. We defined the two-boson correlator $P_{\mathrm \zeta-boson}({\bf k})$ to study the possibility of a phase of paired Cooper pairs, and $\zeta = s,\ d_{xy},\ d^{(1)}_{x^2-y^2},\ d^{(2)}_{x^2-y^2} $.
We explored the existence of boson correlation, and bosons can be paired with both $s$-wave, $d_{xy}$-wave, $d^{(1)}_{x^2-y^2}$-wave and $d^{(2)}_{x^2-y^2}$-wave. In Fig.\ \ref{figS4}(a)-(d), we give the CPQMC simulation results of two-boson correlator $P_{\mathrm \zeta-boson}({\bf k})$ in momentum space for different $t^{\prime}$ with anisotropy $\alpha=0.05$ at $n\sim 0.65$, $\zeta = s,\ d_{xy},\ d^{(1)}_{x^2-y^2},\ d^{(2)}_{x^2-y^2} $, respectively.

\begin{figure*}[ht!]
    \centering
    \includegraphics[width=0.8\linewidth]{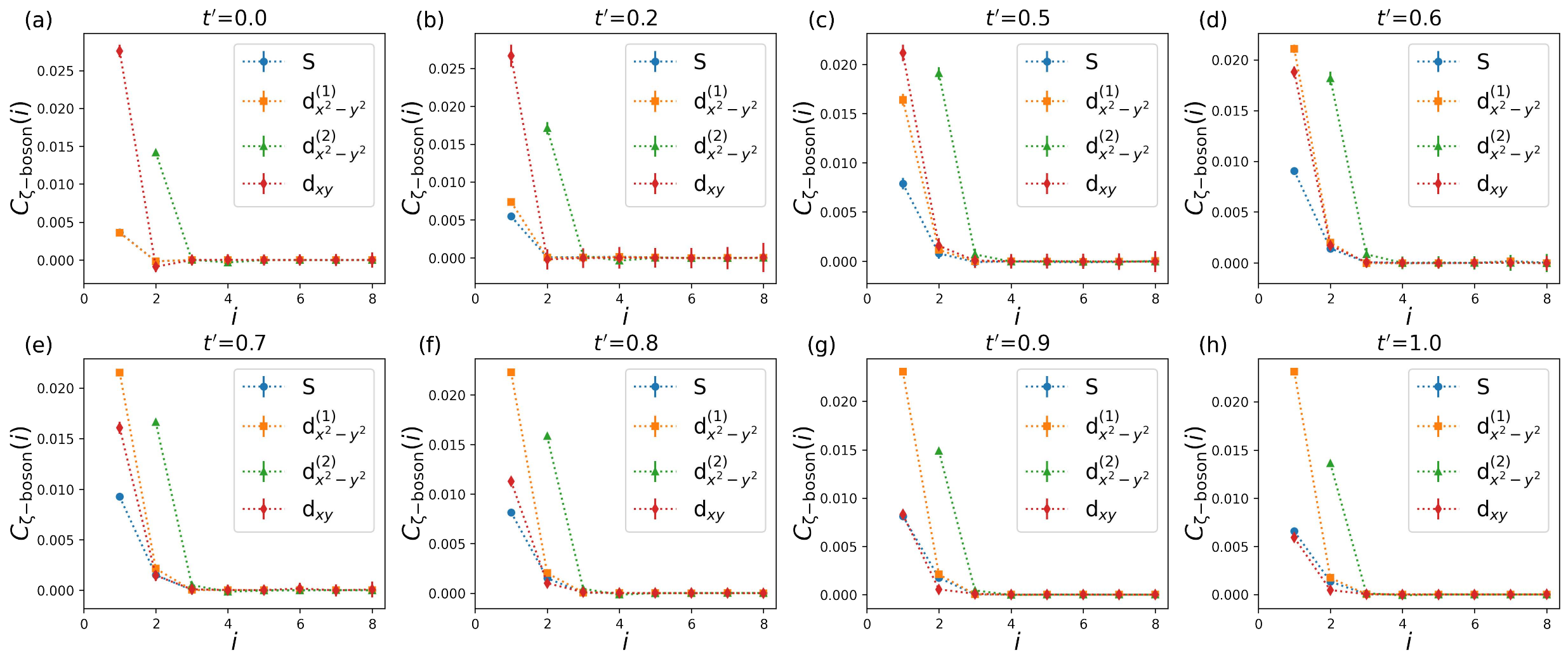}
    \caption{(Color online) The boson pairing correlation $C_{\mathrm \zeta-boson}$ of different boson pairing modes in real space, with $\alpha =0.05$ at $n\sim 0.65$ for $t^{\prime}=$ 0.0, 0.2, 0.5, 0.6, 0.7, 0.8, 0.9, 1.0, and $\zeta = s,\  d^{(1)}_{x^2-y^2},\ d^{(2)}_{x^2-y^2},\ d_{xy}$.  }
    \label{figS5}
\end{figure*}

We extract the maximum values of four pairing modes of boson paired correlation and show in the main text Fig. 3, the variation of boson correlation with the strength of NNN hopping $t^{\prime}$. As $t^{\prime}$ increases, the intensity of the $d^{(1)}_{x^2-y^2}$-wave boson correlation increases and the $d^{(2)}_{x^2-y^2}$-wave does not change much, the proportion of the $d_{xy}$-wave boson correlation decreases, while $s$-wave remains relatively stable with relatively small values. Overall, the primary competition arises between $d_{xy}$-wave and two types of $d_{x^2-y^2}$-wave, and ultimately the $d_{x^2-y^2}$-wave dominates when $t^{\prime} > 0.7$. Among them, the $d^{(1)}_{x^2-y^2}$-wave boson correlation has the greatest enhancement, while the overall change in the $d^{(2)}_{x^2-y^2}$-wave is not significant. This indicates that the change of $t^{\prime}$ has the greatest impact on the $d^{(1)}_{x^2-y^2}$-wave boson correlation.

In Fig.\ \ref{figS5}, we compare the correlations between boson pairs in real space at different $t^{\prime}$. They all decay rapidly to zero, indicating that boson correlation are short-range. In CPBM phase, when $t^{\prime}$ is relatively small, the strength of the $d_{xy}$-wave boson correlation is the highest, however, two types of $d_{x^2-y^2}$-wave dominate when $t^{\prime}$ is large, especially the $d^{(1)}_{x^2-y^2}$-wave. Overall, we suggest its predominant boson pairing mode is always $d$-wave character in Bose metal phase, the system tends toward $d_{x^2-y^2}$ correlation between Cooper pairs when $t^{\prime}$ large, and the system tends towards a $d_{xy}$-wave boson correlation when $t^{\prime}$ is small. Particularly, our CPQMC results indicate that the NNN hopping $t^{\prime}$ being a key tuning parameter to adjust the intensity of boson correlation in the boson metallic phase. This exotic phase of boson-paired states can be termed the $d$-wave Bose Metal.

\section{Competition between $d$-wave boson correlation when $t<t^{\prime}$}

\begin{figure}[htb!]
    \centering
    \includegraphics[width=0.6\linewidth]{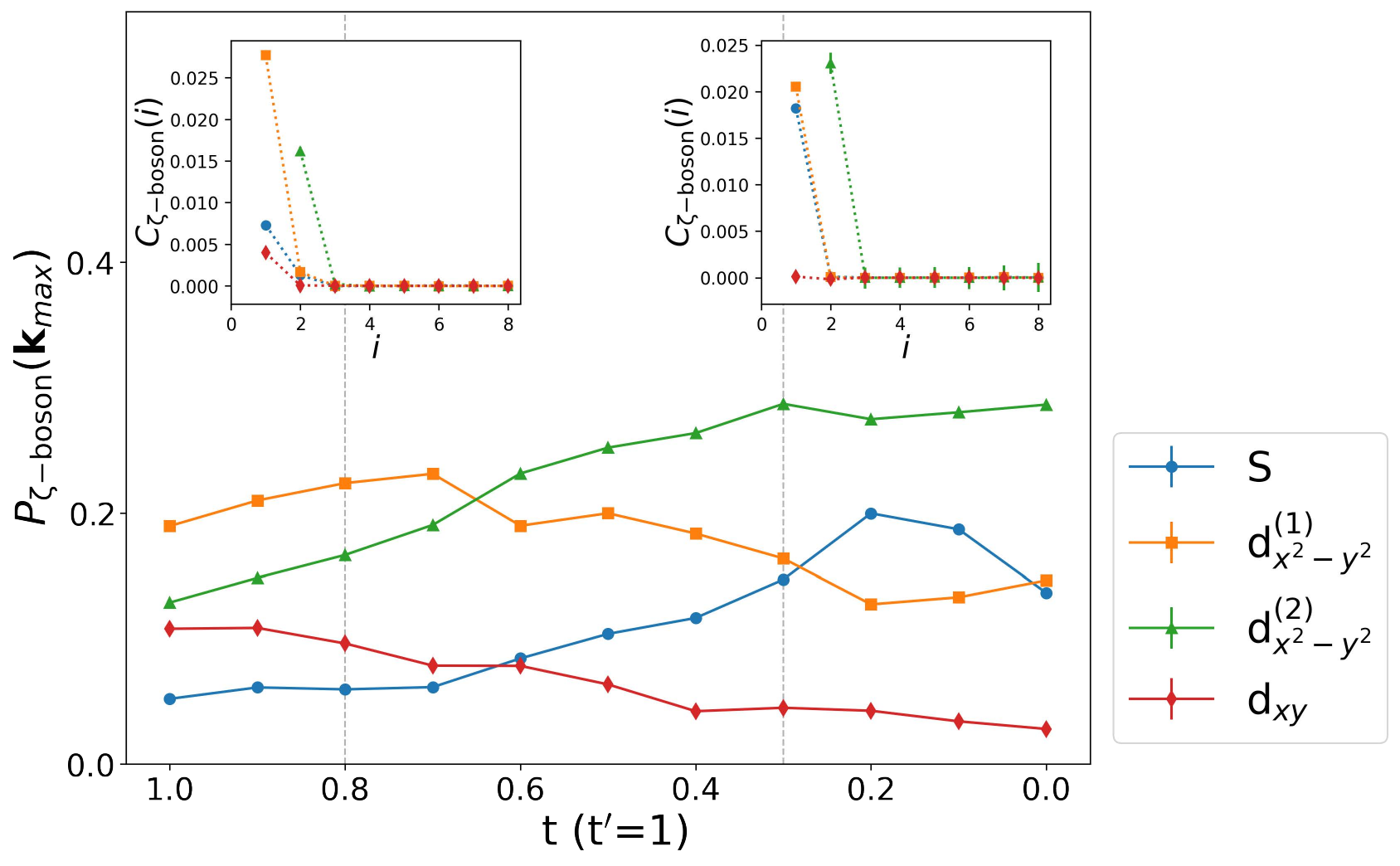}
    \caption{(Color online) Competition between $d$-wave boson correlation for different $t$ when $t<t^{\prime}$. The strength of four pairing modes of bosons paired correlation at $\alpha =0.05$ with $n \sim0.65$ varies by different $t$, and fix $t^{\prime}=1$. The inset shows boson pairing correlations in the real space for $t= 0.8$ and $0.3$, which marked by black dashed lines.  }
    \label{figS6}
\end{figure}

In this section, we will explore a limit case where the NNN hopping $t^{\prime}$ greater than the NN hopping $t$. In this case, we assume fix $t^{\prime}=1$ and $U=-3$ to ensure $U/t^{\prime}=-3$ is compared to $U/t =-3$ in previous discussion. 

In the highly anisotropic regime at filling n$\sim$0.65 for various $t$ values, we also observe the presence of Bose surface in Cooper-pair distribution function. Here, we mainly focus on the correlation between bosons in the Bose metal phase.
We demonstrated changes in the strength of four pairing modes of boson paired correlation by change the NN hopping $t$ (Fig.\ \ref{figS6}), we fix $t^{\prime}=1$ and $\alpha =0.05$ at $n\sim 0.65$. As $t$ decreases, the intensity of the $d^{(2)}_{x^2-y^2}$-wave boson correlation increases and the $d^{(1)}_{x^2-y^2}$-wave does not change much, the proportion of the $d_{xy}$-wave boson correlation has been decreasing. When $t$ is large, the $d^{(1)}_{x^2-y^2}$-wave boson correlation dominates and ultimately the $d^{(2)}_{x^2-y^2}$-wave dominates when $t<0.7$. Moreover, the proportion of the $s$-wave boson correlation increases with the decrease of $t$.

In the inset of Fig.\ \ref{figS6}, we compare the correlations between four pairing modes of boson pairs in real space for $t = 0.8$ and $0.3$. When $t=0.8$, the strength of the $d^{(1)}_{x^2-y^2}$-wave boson correlation is the highest, however, the $d^{(2)}_{x^2-y^2}$-wave dominate when $t=0.3$. Overall, we suggesting its predominant boson pairing mode is always $d$-wave character in Bose metal phase. Particularly, our CPQMC results indicate that the system tends toward $d^{(2)}_{x^2-y^2}$-wave correlation between Cooper pairs when $t$ is small ($t<0.7$).

\section{Incommensurate density wave}

\begin{figure*}[htb!]
    \centering
    \includegraphics[width=1\linewidth]{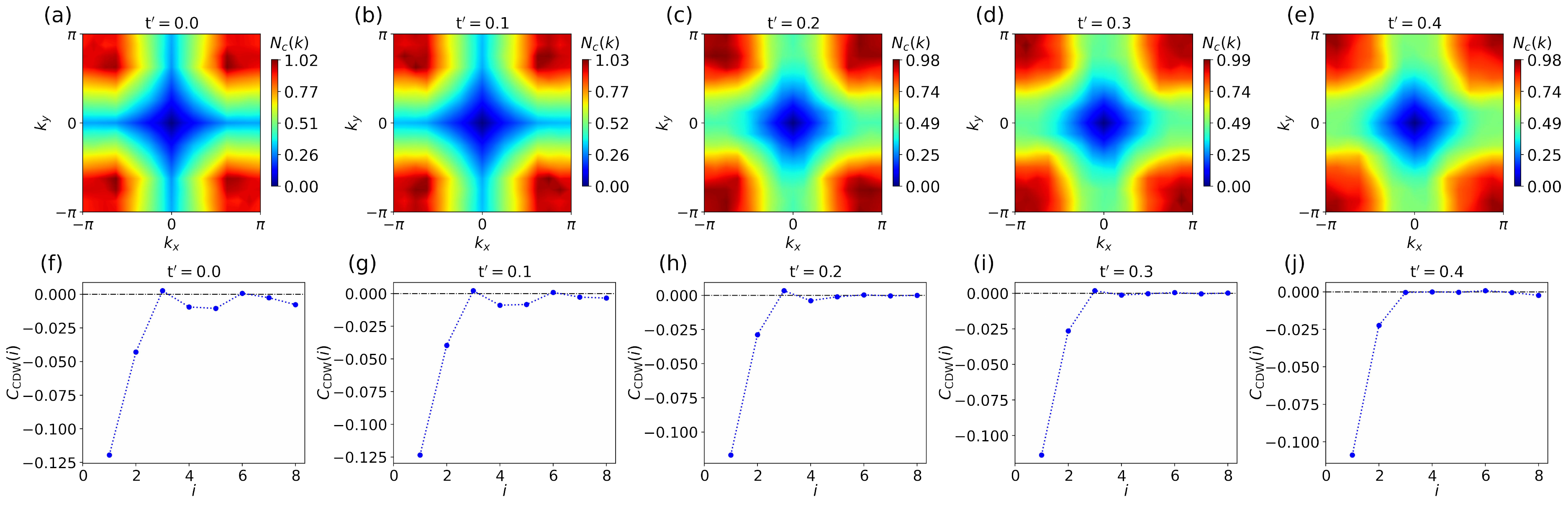}
    \caption{(Color online) Incommensurate density wave. (a)-(e) The charge structure factor for $\alpha =0.05$ with different $t^{\prime}=0.0,\ 0.1,\ 0.2,\ 0.3,\ 0.4$ at $n \sim0.65$ on 16 $\times$ 16 lattice. As $t^{\prime}$ increasing, IDW condenses at non zero momentum point $\bf Q$ = ($2k_F$, $2k_F$), gradually diffusing to the entire momentum space, IDW disappears. (f)-(j) The correlation of charge density wave $C_{\rm CDW}$ in real space for $\alpha =0.05$ with different $t^{\prime}=0.0,\ 0.1,\ 0.2,\ 0.3,\ 0.4$. }
    \label{figS7}
\end{figure*}

We  provide the charge structure factor $N_{\mathrm c}({\bf k})$ for different $t^{\prime} = 0.0,\ 0.1,\ 0.2,\ 0.3,\ 0.4$ in $\alpha=0.05$ at $n\sim 0.65$ in Fig.\ \ref{figS7}(a)-(e), having singular features condenses at non zero momentum point $\bf Q$ = ($2k_F$, $2k_F$) when $t^{\prime}<0.3$. Fermi momentum $k_F$, which is defined as the highest momentum with occupied fermions, the density correlation is expected to oscillate with $2k_F$ wave vector in IDW phase. But when $t^{\prime}$ is large enough ($t^{\prime}>0.3$), the density wave condensed at $\bf Q$ = ($\pi$, $\pi$), IDW naturally converts to weaker CDW in extremely anisotropic region. We observed that the position of condensation is affected by $t^{\prime}$ and the position also moves towards to ($\pi$, $\pi$) point as $t^{\prime}$ increases, which is very similar to change the filling $n$, but the strength of condensation is not related to $t^{\prime}$. We can clearly see that density waves condensed at $\bf Q$ = ($2k_F$, $2k_F$) points in strong anisotropy ($\alpha = 0.05$), and when $t^{\prime}$ increase, the condensated point slowly move to the periphery of the entire Brillouin zone, but there was no significant change in the maximum value of the condensated point. This indicates that the strength of density wave is not closely related to $t^{\prime}$. 

We show the density correlation in real space for different $t^{\prime}= 0.0,\ 0.1,\ 0.2,\ 0.3,\ 0.4$ in Fig.\ \ref{figS7}(f)-(j) consistent with Fig.\ \ref{figS7}(a)-(e). IDW have a non integer multiple of the lattice constant periodic modulation, and it is long-range order of periodic modulation of electron density. Especially, When $t^{\prime} = 0.4$, the density wave have no obvious positive values in real space. In other words, there is no obvious periodic modulation. Although it condenses at the $\bf Q$ = ($\pi$, $\pi$) in momentum space, it cannot be called CDW, we called weaker CDW or delusive CDW. The boundary points of the IDW region in phase diagram are determined comprehensively by whether the density waves condensed at $\bf Q$ = ($2k_F$, $2k_F$) point with a sharp peak in momentum space and whether have a non integer multiple of the lattice constant periodic modulation. We demonstrate that IDW disappears when $t^{\prime}>0.4$ in phase diagram, suggesting that too large $t^{\prime}$ is not conducive to the formation of IDW.

\section{Constraint path quantum Monte Carlo}

The CPQMC method is a quantum Monte Carlo method with a constraint path approximation~\cite{Zhang1995-hn,Zhang1997-mr}. CPQMC method prevents the infamous sign problem \cite{Loh1990-iq} encountered in the DQMC method when dealing with systems that are far from half filling. In CPQMC, the ground state wave function $\psi^{(n)}$ is represented by a finite ensemble of Slater determinants, i.e.,  
$\left|\psi^{(n)}\right\rangle \propto \sum_k\left|\phi_k^{(n)}\right\rangle$,
where $k$ is the index of the Slater determinants, and $n$ is the number of iteration. The overall normalization factor of the wave function has been omitted here. The propagation of the Slater determinants dominates the computational time, as follows
\begin{equation}
\left|\phi_k^{(n+1)}\right\rangle \leftarrow \int d \vec{x} P(\vec{x}) B(\vec{x})\left|\phi_k^{(n)}\right\rangle .
\label{propagation}
\end{equation}
where $\vec{x}$ is the auxiliary-field configuration, that we select according to the probability distribution function $P(\vec{x})$. The propagation includes the matrix multiplication of the propagator $B(\vec{x})$ and $\phi_k^{(n)}$. 
% The trial wavefunction $\phi^{(0)}$ is used to initialize the random walkers. 
% After a series of equilibrium steps, the walkers are the Monte Carlo samples of the ground state wavefunction, so we could obtain the ground state property through measurements. 
After a series of equilibrium steps, the walkers are the Monte Carlo samples of the ground state wave function $\phi^{(0)}$ and ground-state properties can be measured. 

The random walk formulation suffers from the sign problem because of the fundamental symmetry existing between the fermion ground state $\left|\psi_0\right\rangle$ and its negative $-\left|\psi_0\right\rangle $. 
In more general cases, walkers can cross $\mathcal{N}$ in their propagation by $e^{-\Delta \tau H}$
whose bounding surface $\mathcal{N}$ is defined by $\left\langle\psi_0 \mid \phi\right\rangle=0$ and is in general $unknown$.
Once a random walker reaches $\mathcal{N}$, it will make no further contribution to the representation of the ground state since
\begin{equation}
\left\langle\psi_0 \mid \phi\right\rangle=0 \Rightarrow\left\langle\psi_0\left|e^{-\tau H}\right| \phi\right\rangle=0 \quad \text { for any } \tau .
\label{no contribution}
\end{equation}

Paths that result from such a walker have equal probability of being in either half of the Slater-determinant space. Computed analytically, they would cancel, but without any knowledge of $\mathcal{N}$, they continue to be sampled in the random walk and become Monte Carlo noise.

The decay of the signal-to-noise ratio, i.e., the decay of the average sign of $\left\langle\psi_T \mid \phi\right\rangle$, occurs at an exponential rate with imaginary time.
To eliminate the decay of the signal-to-noise ratio, we impose the constrained path approximation. It requires that each random walker at each step has a positive overlap with the trial wave function $\left|\psi_T\right\rangle$ :

\begin{equation}
\left\langle\psi_T \mid \phi_k^{(n)}\right\rangle>0 .
\label{constrained path approximation}
\end{equation}

This yields an approximate solution to the ground-state wave function, $\left|\psi_0^c\right\rangle=\Sigma_\phi|\phi\rangle$, in which all Slater determinants $|\phi\rangle$ satisfy Eq.~\ref{no contribution}. From Eq.~\ref{constrained path approximation}, it follows that this approximation becomes exact for an exact trial wave function $\left|\psi_T\right\rangle=\left|\psi_0\right\rangle $. In our previous work \cite{Cao2024-qd}, we find that both numerically exact DQMC and sign-constraint CPQMC gives fingerprint of CPBM (nonzero peaks in $N_{s-pair}$) in the quasi-1D lattice of two-leg ladder. We also provide the choice of hyperparameters in CPQMC \cite{Cao2024-qd}.  This could support the applicability of using CPQMC in studying CPBM in 2D lattice.

\bibliography{ref}